\documentclass[preprint,3p,12pt]{elsarticle}

\usepackage{amssymb}
\usepackage{latexsym,amsmath,subfigure}

\journal{}

\newtheorem{thm}{Theorem}[section]

\newtheorem{rem}[thm]{Remark}
\newcommand{\eps}{\varepsilon}
\newcommand{\abs}[1]{\left\vert#1\right\vert}
\newcommand{\set}[1]{\left\{#1\right\}}

\newcommand{\p}{\partial}

\begin{document}

\begin{frontmatter}



\title{Determination of the unknown support of thin penetrable inclusions within a half-space from multi-frequency response matrix}

\author[KMU]{Won-Kwang Park}
\ead{parkwk@kookmin.ac.kr}
\author[Supelec]{Dominique Lesselier\corref{lesselier}}
\ead{dominique.lesselier@lss.supelec.fr}
\cortext[lesselier]{Corresponding author.}

\address[KMU]{Department of Mathematics, Kookmin University, Seoul, 136-702, Korea.}
\address[Supelec]{D\'epartement de Recherche en \'Electromagn\'etisme - Laboratoire des Signaux et Syst\`{e}mes, UMR8506 (CNRS-Sup\'elec-Universit\'e Paris-Sud~11) 91192 Gif-sur-Yvette cedex, France.}

\begin{abstract}
Thin, penetrable electromagnetic inclusions buried within a half space are retrieved from Multi-Static Response (MSR) matrix data collected above this half space at several frequencies. A non-iterative algorithm is proposed to that aim. It relies on the modeling of the MSR matrix according to a rigorous asymptotic expansion of the scattering amplitude. The supporting curves of single as well as multiple thin inclusions are retrieved in fairly good fashion. This is illustrated by a set of numerical inversions that are carried out from synthetic data, both noiseless and noisy ones.
\end{abstract}


\end{frontmatter}




\section{Introduction}\label{Sec1}
To achieve a reliable imaging of thin inclusions and cracks still remains a challenging problem in the area of electromagnetic non-destructive evaluation. This is true even if simple models of those defects are chosen, e.g., each is assimilated with a thin (with respect to the wave length of the probing electromagnetic field), homogeneous penetrable layer which extends around a properly smooth supporting curve, transmission conditions being satisfied at its boundary, with the limiting case of impenetrability being taken care of as well. Yet, ill-posedness and inherent nonlinearity of the inverse problem at hand often impair that achievement.

So, over a number of decades, many imaging methods have been suggested. Almost all are based on least-square minimization iterative schemes, so consequently, initial guess close enough to the unknown defect and suitable regularization tuned to the problem at hand are essential. Without, one might suffer from the occurrence of several minima, large computational costs, and so on.

Recently, MUSIC (MUltiple SIgnal Classification)-type, non-iterative algorithms have been developed in order to overcome some of these difficulties. They aim at the retrieval of the location and shape of penetrable, electromagnetically thin inclusions as well as the one of perfectly conducting ones at a fixed single frequency. From several results in \cite{PL1,PL2,PL3}, such MUSIC algorithms appear both fast and robust, while easily extending to several disjoint inclusions. However, whenever applied to limited view data (one cannot illuminate the unknown inclusions from all around neither collect the scattered fields all around), for example, when those inclusions are embedded within a half space, emitters and receivers being set above it, retrievals are much poorer, as is seen in \cite{AIL,P1,P2,PL3} for either volumetric inclusions or thin ones. Yet, recently, an effective and robust multi-frequency algorithm has been proposed so as to find the location of small cracks  \cite{AGKPS} but a suitable algorithm for extended ones appears still in need.

The purpose of the present paper is to propose an effective, non-iterative imaging algorithm which is able to work on limited view data in order to retrieve electromagnetically penetrable thin inclusions buried within a homogeneous (lower) half space. It is based on the fact that the Multi-Static Response (MSR) matrix which one can collect in that setting (at least samples of it) can be modeled via a rigorously derived asymptotic expansion formula of the scattering amplitude in the presence of the inclusions. A number of numerical simulations will then illustrate how the proposed imaging algorithm operated at several frequencies behaves, and enhances the performance of imaging that would be based on the statistical hypothesis testing, refer to \cite{AGKPS}.

The paper is structured as follows. In section \ref{Sec2}, the two-dimensional direct scattering problem is sketched and the asymptotic formula for the scattering amplitude are introduced. In section \ref{Sec4}, the multi-frequency based non-iterative imaging algorithm is outlined. In section \ref{Sec5}, the numerical results are shown. A short conclusion follows (section \ref{Sec6}).

\section{Direct scattering problem}\label{Sec2}
In this section, we briefly discuss the two-dimensional, time-harmonic electromagnetic scattering from a thin inclusion buried in a homogeneous half space. A more detailed description is found in \cite{AIL}.

Let us define the lower half space and the upper half space represented as
\[\mathbb{R}_{-}^{2}=\set{(x_1,x_2)^T\in\mathbb{R}^2:x_2<0}\quad\mbox{and}\quad\mathbb{R}_{+}^{2}=\set{(x_1,x_2)^T\in\mathbb{R}^2:x_2>0},\]
respectively.

\begin{figure}
\begin{center}
\includegraphics[width=0.25\textwidth,keepaspectratio=true,angle=0]{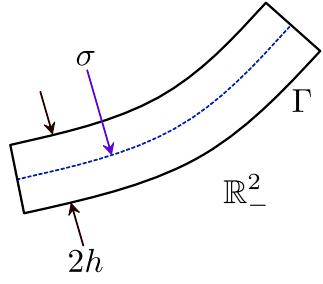}
\caption{\label{ThinInclusion}Sketch of the thin inclusion $\Gamma$ buried within a half-space $\mathbb{R}_{-}^2$.}
\end{center}
\end{figure}

A thin penetrable inclusion, $\Gamma$ is fully embedded in the homogeneous space $\mathbb{R}_{-}^{2}$. It is curve-like, i.e., it is localized in the neighborhood of a curve:
\[\Gamma=\set{x+\eta n(x):x\in\sigma,~\eta\in(-h,h)},\]
where $\sigma$ is a simple, smooth curve contained in $\mathbb{R}_{-}^{2}$, with strictly positive distance from the boundary, $\p\mathbb{R}^{2}=\set{(x_1,x_2)^T\in\mathbb{R}^2:x_2=0}$, $n(x)$ is a unit normal to $\sigma$ at $x$, and $h$ is a small (with respect to the wavelength of the electromagnetic field in the embedding space at the given frequency of operation $\omega$), positive constant which specifies the thickness of the inclusion, refer to Fig. \ref{ThinInclusion}.

All materials are characterized by their dielectric permittivity and magnetic permeability at $\omega$; $\eps_-$, $\eps_+$ and $\eps_T$ denote the electric permittivity of $\mathbb{R}_{-}^{2}$, $\mathbb{R}_{+}^{2}$ and $\Gamma$, respectively.The magnetic permeability can be denoted likewise. So, the piecewise-constant electric permittivity $0<\eps(x)<+\infty$ and magnetic permeability $0<\mu(x)<+\infty$ as
\[\eps(x)=\left\{\begin{array}{ccl}
\eps_{+}&\mbox{for}&x\in\mathbb{R}_{+}^{2}\\
\eps_{-}&\mbox{for}&x\in\mathbb{R}_{-}^{2}\backslash\overline{\Gamma}\\
\eps_T&\mbox{for}&x\in\Gamma
\end{array}\right.
\quad\mbox{and}\quad
\mu(x)=\left\{\begin{array}{ccl}
\mu_{+}&\mbox{for}&x\in\mathbb{R}_{+}^{2}\\
\mu_{-}&\mbox{for}&x\in\mathbb{R}_{-}^{2}\backslash\overline{\Gamma}\\
\mu_T&\mbox{for}&x\in\Gamma.
\end{array}\right.\]
For convenience, the electric permittivity $\eps_0(x)$ and magnetic permeability $\mu_0(x)$ when there are no inclusions are such as
\[\eps_0(x)=\left\{\begin{array}{ccl}
\eps_{+}&\mbox{for}&x\in\mathbb{R}_{+}^{2}\\
\eps_{-}&\mbox{for}&x\in\mathbb{R}_{-}^{2}
\end{array}\right.
\quad\mbox{and}\quad
\mu_0(x)=\left\{\begin{array}{ccl}
\mu_{+}&\mbox{for}&x\in\mathbb{R}_{+}^{2}\\
\mu_{-}&\mbox{for}&x\in\mathbb{R}_{-}^{2}
\end{array}\right.\]
Accordingly, the piecewise positive real-valued wavenumber $k(x)$ when there are no inclusions can be written as
\[k(x)=\left\{\begin{array}{ccl}
k_{+}=\omega\sqrt{\eps_+\mu_+}&\mbox{for}&x\in\mathbb{R}_{+}^{2}\\
k_{-}=\omega\sqrt{\eps_-\mu_-}&\mbox{for}&x\in\mathbb{R}_{-}^{2}.
\end{array}\right.\]

Let now $\theta=(\theta_1,\theta_2)$ be a two-dimensional vector on the unit circle $S^1\subset\mathbb{R}^2$ and $u_i(x)=e^{ik_+\theta\cdot x}$ be a planar incident wavefield generated in the upper half space on $\mathbb{R}_{+}^{2}$. At $\omega$, let $u(x)$ denote the total field which satisfies the Helmholtz equation
\[\nabla\cdot\left(\frac{1}{\mu(x)}u(x)\right)+\omega^2 \eps(x)u(x)=0\quad\mbox{in}\quad\mathbb{R}^2,\]
transmission conditions holding at boundaries $\p\mathbb{R}^2$ and $\p\Gamma$.

Let $u_0(x)$ be the plane-wave solution to the Helmholtz equation in the absence of inclusions. Then, within each half space, $u_s(x)=u(x)-u_0(x)$ is required to satisfy the Sommerfeld radiation condition
\[\lim_{|x|\to\infty}\sqrt{|x|}\left(\frac{\p u_s(x)}{\p|x|}-ik(x)u_s(x)\right)=0\]
uniformly into all directions $\hat{x}=\frac{x}{|x|}$.

The scattering amplitude is defined as a function $K(\hat{y},\theta)$ that satisfies
\[u_s(y)=\frac{e^{ik|y|}}{\sqrt{|y|}}K(\hat{y},\theta)+o\left(\frac{1}{\sqrt{|y|}}\right)\]
as $|y|\longrightarrow\infty$ uniformly on $\hat{y}=\frac{y}{|y|}$ and $\theta\in S^1$.

In order to express the asymptotic formula for $K(\hat{y},\theta)$, let us denote $v\in\mathbb{C}^2$ and $T(\hat{x})\in\mathbb{C}$ as
\[v(\hat{x})=\left(\xi\hat{x}_1,\mbox{sign}(\hat{x}_2)\sqrt{1-\xi^2\hat{x}_1^2}\right)^T,\quad
T(\hat{x})=\frac{2\mu_-\xi\hat{x}_2}{\mu_-\xi\hat{x}_2+\mu_+\mbox{sign}(\hat{x}_2)\sqrt{1-\xi^2\hat{x}_1^2}}\]
and a positive, symmetric matrix $\mathcal{A}(x)$ satisfying (see \cite{AK,BF,PL2})
\begin{itemize}
\item $\mathcal{A}(x)$ has eigenvectors $\tau(x)$ and $n(x)$
\item The eigenvalue corresponding to $\tau(x)$ is $\displaystyle 2\mu_-\left(\frac{1}{\mu_T}-\frac{1}{\mu_-}\right)$
\item The eigenvalue corresponding to $n(x)$ is $\displaystyle 2\mu_-\left(\frac{1}{\mu_-}-\frac{\mu_T}{\mu_-^2}\right)$
\end{itemize}
where $\tau(x)$ and $n(x)$ are unit vectors respectively tangential and normal to $x\in\sigma$.

Let us emphasize that in the problem at hand, the measurement of incident and observed fields is restricted to the upper half space $\mathbb{R}_{+}^2$. For convenience, we divide the unit circle $S^1$ into
\[S_{+}^1=\set{x\in\mathbb{R}_{+}^2:|x|=1}\quad\mbox{and}\quad S_{-}^1=\set{x\in\mathbb{R}_{-}^2:|x|=1}.\]
In that configuration, and using material from \cite{AIL,AK,BF,BFV,CV}, we obtain the following result.

\begin{thm}For every $\hat{y}\in S_{+}^1$ and $\theta\in S_{-}^1$, the asymptotic formula for the scattering amplitude $K(\hat{y},\theta)$ is expressed as
\begin{align}\label{SAK}
\begin{aligned}
K(\hat{y},\theta)=&h\frac{(k_-)^2\mu_+(1+i)}{4\mu_-\sqrt{k_+\pi}}T(\hat{y})T(\theta)\bigg[\int_{\sigma}\left(\frac{\eps_T}{\eps_-}-1\right)e^{-ik_-(v(\hat{y})-v(\theta))\cdot x}d\sigma(x)\\
&-\int_{\sigma}v(\hat{y})\cdot\mathcal{A}(x)\cdot v(\theta)e^{-ik_-(v(\hat{y})-v(\theta))\cdot x}d\sigma(x)\bigg]+o(h),
\end{aligned}
\end{align}
where the remaining term $o(h)$ is independent of points $x\in\sigma$.
\end{thm}
Let us notice that the asymptotic expansion (\ref{SAK}) is not obtained by mere application of the Born approximation. One should refer to \cite{AIL} for a similar analysis in the case of small volumetric inclusions.

\section{Non-iterative imaging algorithm}\label{Sec4}
In this section, we apply the asymptotic formula for the scattering amplitude (\ref{SAK}) so as to build up the imaging algorithm. To do so, we use the eigenvalue structure of the Multi-Static Response (MSR) matrix $\mathcal{K}=(K_{jl})$, whose element $K_{jl}$ is the amplitude collected at observation number $j$ for the incident wave numbered $l$.

Let us assume that for a given frequency $\omega$, the thin inclusion is divided into $M$ different segments of size of order $\frac{\lambda}{2}$. Having in mind the Rayleigh resolution limit from far-field data, any detail less than one-half of the wavelength cannot be retrieved, and only one point, say $x_m$ for $m=1,2,\cdots,M$, at each segment is expected to contribute to the image space of the response matrix $\mathcal{K}$, refer to \cite{A,ABC,AKLP,PL1,PL2,PL3}.

For simplicity, let us denote the constant $C=h\frac{(k_-)^2\mu_+(1+i)}{4\mu_-\sqrt{k_+\pi}}$ and remove the residue term $o(h)$ from (\ref{SAK}). Then, for each $\hat{y}_j=-\theta_j$, the $jl-$th element of the MSR matrix $K_{jl}\in\mathbb{C}$, $j,l=1,2,\cdots,N$, is
\begin{align}
\begin{aligned}\label{MSR}
K_{jl}=&CK(\hat{y}_j,\theta_l)\bigg{|}_{\hat{y}_j=-\theta_j}\\
=&CT(\hat{y}_j)T(\theta_l)\int_{\sigma}\bigg[\left(\frac{\eps_T}{\eps_-}-1\right)+v(\theta_j)\cdot\mathcal{A}(x)\cdot v(\theta_l)\bigg]e^{i k_-(v(\theta_j)+v(\theta_l))\cdot x}d\sigma(x)\\
\approx&CT(\theta_j)T(\theta_l)\frac{\abs{\sigma}}{M}\sum_{m=1}^{M}\bigg{[}\left(\frac{\eps_T}{\eps_-}-1\right)+2\bigg(\frac{\mu_-}{\mu_T}-1\bigg)v(\theta_j)\cdot\tau(x_m)v(\theta_l)\cdot\tau(x_m)\\
&+2\bigg(1-\frac{\mu_T}{\mu_-}\bigg)v(\theta_j)\cdot n(x_m)v(\theta_l)\cdot n(x_m)\bigg{]}e^{ik_-(v(\theta_j)+v(\theta_l))\cdot x_m},
\end{aligned}
\end{align}
where $|\sigma|$ denotes the length of $\sigma$.

Notice that MSR matrix $\mathcal{K}$ can be decomposed as follows:
\begin{equation}\label{EVD}
\mathcal{K}=\mathcal{D}\mathcal{E}\mathcal{D}^T.
\end{equation}
Here, the matrix $\mathcal{E}\in\mathbb{R}^{3M\times3M}$ is a diagonal matrix with component
\[\mathcal{E}=h\frac{(k_-)^2\mu_+(1+i)}{4\mu_-\sqrt{k_+\pi}}\frac{|\sigma|}{M}\left(\begin{array}{cc}E_{\eps}&0\\0&E_{\mu}\end{array}\right),\]
where
\begin{align*}
E_{\eps}&=M\times M\mbox{ diagonal matrix with components }\left(\frac{\eps_T}{\eps_-}-1\right),\\
E_{\mu}&=2M\times 2M \mbox{ diagonal matrix with }2\times2\mbox{ blocks }\left(\begin{array}{cc}2\left(\frac{\mu_-}{\mu_T}-1\right)&0\\0&2\left(1-\frac{\mu_T}{\mu_-}\right)\end{array}\right)
\end{align*}
and the matrix $\mathcal{D}\in\mathbb{C}^{N\times3M}$ can be written as follows
\[\left[D_\eps^1\quad D_\eps^2\quad \cdots\quad D_\eps^M\quad D_\mu^1\quad D_\mu^2\quad \cdots\quad D_\mu^{2M}\right]\]
where
\begin{align*}
D_{\eps}^m&=\left(T(\theta_1)e^{ik_- v(\theta_1)\cdot x_m},T(\theta_2)e^{ik_- v(\theta_2)\cdot x_m},\cdots,T(\theta_N)e^{ik_- v(\theta_N)\cdot x_m}\right)^T,\\
D_{\mu}^{2(m-1)+s}&=\left(\xi_s(x_m)\cdot v(\theta_1)T(\theta_1)e^{ik_- v(\theta_1)\cdot x_m},\cdots,\xi_s(x_m)\cdot v(\theta_N)T(\theta_N)e^{ik_- v(\theta_N)\cdot x_m}\right)^T
\end{align*}
with
\[\xi_s(x_m):=\left\{\begin{array}{rcl}\tau(x_m)&\mbox{if}&s=1\\n(x_m)&\mbox{if}&s=2.\\\end{array}\right.\]

Let us notice that with the representation (\ref{EVD}), $\mathcal{K}$ is symmetric but not Hermitian (a Hermitian matrix could be formed as $\mathcal{K}\overline{\mathcal{K}}$). Since $\mathcal{K}$ is is not self-adjoint, a Singular Value Decomposition (SVD) has to be used instead of the eigenvalue decomposition. Let us perform this decomposition of matrix $\mathcal{K}$ and let $M$ be the number of nonzero singular values for the given $\omega$. Then, $\mathcal{K}$ can be represented as follows:
\[\mathcal{K}=\mathcal{U}(\omega)\mathcal{S}(\omega)\overline{\mathcal{V}(\omega)}^T\approx\sum_{m=1}^{M}u_m(\omega)s_m(\omega)\overline{v}_m^T(\omega),\]
where $s_m(\omega)$ are the singular values, $u_m(\omega)$ and $v_m(\omega)$ are the left and right singular vectors of $\mathcal{K}$ for $m=1,2,\cdots,M$.

Based on the above singular value decomposition, the imaging algorithm is developed as follows. For $c\in\mathbb{R}^3\backslash\set{0}$, let us define a vector
\[d(x;\omega)=\left(c\cdot(1,v(\theta_1))T(\theta_1)e^{ik_-v(\theta_1)\cdot x},\cdots,c\cdot(1,v(\theta_N))T(\theta_N)e^{ik_-v(\theta_N)\cdot x}\right)^T\]
and corresponding normalized vector $\hat{d}(x;\omega)=\frac{d(x;\omega)}{||d(x;\omega)||}$. Then for $m=1,2,\cdots,M$,
\begin{equation}\label{Approx1}
u_m\sim e^{i\gamma_1}\hat{d}(x_m;\omega)\quad\mbox{and}\quad v_m\sim e^{-i\gamma_2}\overline{\hat{d}(x_m;\omega)}
\end{equation}
with $\gamma_1+\gamma_2=\frac{\pi}{2}$, refer to \cite{HHSZ}. Since the first $M$ columns of the matrix $\mathcal{U}(\omega)$ and $\mathcal{V}(\omega)$, $\set{u_1(\omega),u_2(\omega),\cdots,u_M(\omega)}$ and $\set{v_1(\omega),v_2(\omega),\cdots,v_M(\omega)}$, are orthonormal, one can easily find that
\begin{align}
\begin{aligned}\label{Approx2}
\langle\hat{d}(x;\omega),u_m(\omega)\rangle&\ne 0\quad\mbox{and}\quad\langle\hat{d}(x;\omega),\overline{v}_m(\omega)\rangle\ne 0\quad\mbox{if}\quad x=x_m\\
\langle\hat{d}(x;\omega),u_m(\omega)\rangle&\approx 0\quad\mbox{and}\quad\langle\hat{d}(x;\omega),\overline{v}_m(\omega)\rangle\approx 0\quad\mbox{if}\quad x\ne x_m
\end{aligned}
\end{align}
for $m=1,2,\cdots,M$, where $\langle a,b\rangle=\overline{a}\cdot b$.

Now, let us construct the following image function with MSR matrix at fixed frequency $\omega$:
\begin{equation}\label{ImagingFunctionSingle}
\mathrm{W}_S(x)=\sum_{m=1}^{M}|\langle\hat{d}(x;\omega),u_m(\omega)\rangle\langle\hat{d}(x;\omega),\overline{v}_m(\omega)\rangle|.
\end{equation}
Based on (\ref{Approx1}) and (\ref{Approx2}), the map of $\mathrm{W}_S(x)$ is expected to exhibit peaks of magnitude of $1$ at location $x_m$ for $m=1,2,\cdots,M$ and of small magnitude at $x\in\mathbb{R}_-^2\backslash\overline{\Gamma}$.

Unfortunately, the image functional (\ref{ImagingFunctionSingle}) at single frequency offers an image with poor resolution, refer to \cite{AGKPS,P1,P2}. In order to improve the imaging performance, we suggest a normalized image functional at several frequencies $\{\omega_f:f=1,2,\cdots,F\}$:
\begin{equation}\label{ImagingFunction}
\mathrm{W}_F(x)=\frac{1}{F}\sum_{f=1}^{F}\sum_{m=1}^{M_f}|\langle\hat{d}(x;\omega_f),u_m(\omega_f)\rangle\langle\hat{d}(x;\omega_f),\overline{v}_m(\omega_f)\rangle|,
\end{equation}
where $M_f$ is the number of nonzero singular values of the MSR matrix at $\omega_f$ for $f=1,2,\cdots,F$. Then, similarly with the (\ref{ImagingFunctionSingle}), the map of $W(x)$ should exhibit peaks of magnitude of $1$ at location $x_m$ for $m=1,2,\cdots,M_f$, and of small magnitude at $x\in\mathbb{R}_-^2\backslash\overline{\Gamma}$. A suitable number of $M_f$ for each frequency $\omega_f$ can be found via careful thresholding, see \cite{PL1,PL2,PL3} for instance. It is worth mentioning that multiple frequencies should enhance the imaging performance via higher signal-to-noise ratio (SNR) based on the statistical hypothesis testing, refer to appendix \ref{SecA} (see \cite{AGKPS} also).

\begin{rem}\label{Rem}For effective imaging, to properly choose the vector $c=(a,b_1,b_2)\in\mathbb{R}^3\backslash\set{0}$ is a strong prerequisite. The choice comes from the structure of the elements of the MSR matrix (\ref{MSR}). First, $c=(1,0,0)$ is a good, easy choice for a purely dielectric contrast case. But, for a purely magnetic contrast case, the vector $b$ must be a linear combination of the tangential and normal vectors while keeping $a=0$. To appraise the tangential (and also normal) vector of the supporting curve $\sigma$ should not be straightforward, yet we can obtain a good image with $(b_1,b_2)=(1,0)$ and $(b_1,b_2)=(0,1)$ for $\mu_T<\mu_-$ and $\mu_T>\mu_-$, respectively. For a double contrast case, $c$ should be a combination of the above two results. A detailed discussion can be found in \cite[section 4.3.1]{PL2}.
\end{rem}

\section{Numerical examples}\label{Sec5}
\begin{figure}
\begin{center}
\includegraphics[width=0.4\textwidth,keepaspectratio=true,angle=0]{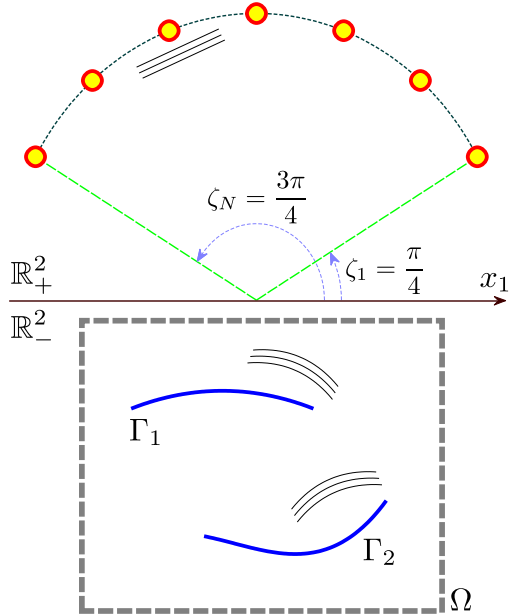}
\caption{\label{Conf}Sketch of the test configuration.}
\end{center}
\end{figure}

In this section, numerical examples are provided. Throughout, the thickness $h$ of the thin inclusion are set to $0.015$ and the applied frequency is $\omega_f=\frac{2\pi}{\lambda_f}$; here $\lambda_f$, $f=1,2,\cdots,F$, is the given wavelength. In this paper, frequencies $\omega_f$ are equi-distributed within the interval $[\omega_1,\omega_F]$. As for the observation directions $\hat{y}_j$, they are taken as
\[\hat{y}_j=-\left(\cos\zeta_j,\sin\zeta_j\right),\quad\zeta_j=\alpha+(\beta-\alpha)\frac{j-1}{N-1},\]
where $\alpha=\frac{\pi}{4}$ and $\beta=\frac{3\pi}{4}$ for $j=1,2,\cdots,N$. See Fig. \ref{Conf} for an illustration of the test configuration.

Note that when the upper half space is more refractive than the lower one, i.e., $k_+>k_-$, the number $N_+$ of propagating transmitted waves might be less than $N$, refer to Table \ref{Configuration1} (see \cite{AIL} also).

Two $\sigma_j$ characteristic of the thin inclusion $\Gamma_j$ are chosen for illustration:
\begin{align}
\begin{aligned}
\sigma_1&=\set{(z-0.2,-0.5z^2-1.5):z\in[-0.5,0.5]}\\
\sigma_2&=\set{(z+0.2,z^3+z^2-2.5):z\in[-0.5,0.5]}\\
\end{aligned}
\end{align}
and define $\Gamma_{\mbox{\tiny M}}=\Gamma_1\cup\Gamma_2$ for multiple inclusions. Throughout this section, we denote $\eps_j$ and $\mu_j$ be the permittivities and permeabilities of $\Gamma_j$, respectively.

It is worth emphasizing that the data set of the MSR matrix $\mathcal{K}$ is computed within the framework of the Foldy-Lax equation, refer to \cite{DMG,PL2,TKDA}. Then, a white Gaussian noise with 20dB signal-to-noise ratio (SNR) is added to the unperturbed data in order to investigate the robustness of the proposed algorithm while at least partly alleviating inverse crime. In order to obtain the number of nonzero singular values $M_f$ at each frequency $\omega_f$, a $0.01$-threshold scheme (choosing first $j$ singular values $s_j(\omega_f)$ such that $\frac{s_j(\omega_f)}{s_1(\omega_f)}\geq0.01$) is adopted. A more detailed discussion of thresholding can be found in \cite{PL1,PL2,PL3}. As for the step size of the search points $x\in\Omega$, it is taken of the order of $0.02$.

A is observed from the numerical experiments led in \cite{AIL}, similar results were obtained for both cases $\eps_+>\eps_-$ and $\eps_+=\eps_-$ (permeability, and both contrast cases also). Thus, two different situations of interest are henceforth considered.

\subsection{Permittivity contrast case: $\eps_T\ne\eps_-$ and $\mu_T=\mu_-=\mu_+$}
\begin{table}
\begin{center}
\begin{tabular}{c|c|c|c|c|c}
\hline inclusion&$N$&$N_+$&$F$&frequency range&search domain\\
\hline\hline
$\Gamma_1$&$32$&$24$&$30$&$\omega\in[\frac{2\pi}{0.4},\frac{2\pi}{0.2}]$&$\Omega=[-1,1]\times[-3,-1]$\\
$\Gamma_2$&$40$&$28$&$36$&$\omega\in[\frac{2\pi}{0.3},\frac{2\pi}{0.1}]$&$\Omega=[-1,1]\times[-3,-1]$\\
$\Gamma_{\mbox{\tiny M}}$&$48$&$32$&$40$&$\omega\in[\frac{2\pi}{0.2},\frac{2\pi}{0.1}]$&$\Omega=[-1,1]\times[-3,-1]$\\
\hline
\end{tabular}
\caption{\label{Configuration1}(Permittivity contrast case) Test configuration for $\Gamma_1$, $\Gamma_2$ and $\Gamma_{\mbox{\tiny M}}$.}
\end{center}
\end{table}

\subsubsection{$\eps_+>\eps_-$ case}
\begin{figure}
\begin{center}
\subfigure[$\Gamma_1$ with $\eps_1=5$]{\label{SingleEps-1a}\includegraphics[width=0.49\textwidth]{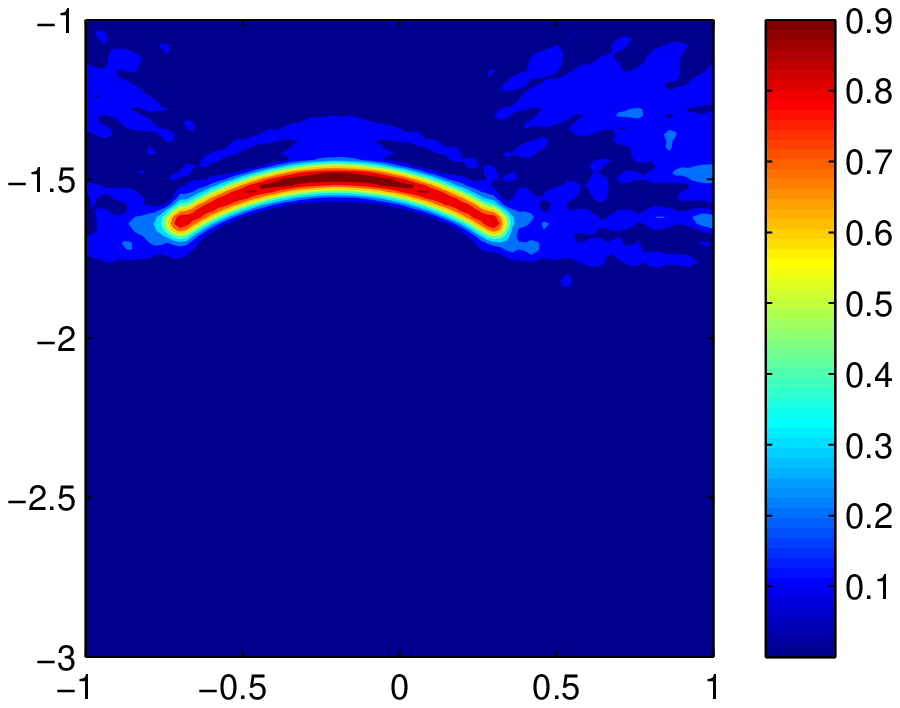}}
\subfigure[$\Gamma_2$ with $\eps_2=5$]{\label{SingleEps-1b}\includegraphics[width=0.49\textwidth]{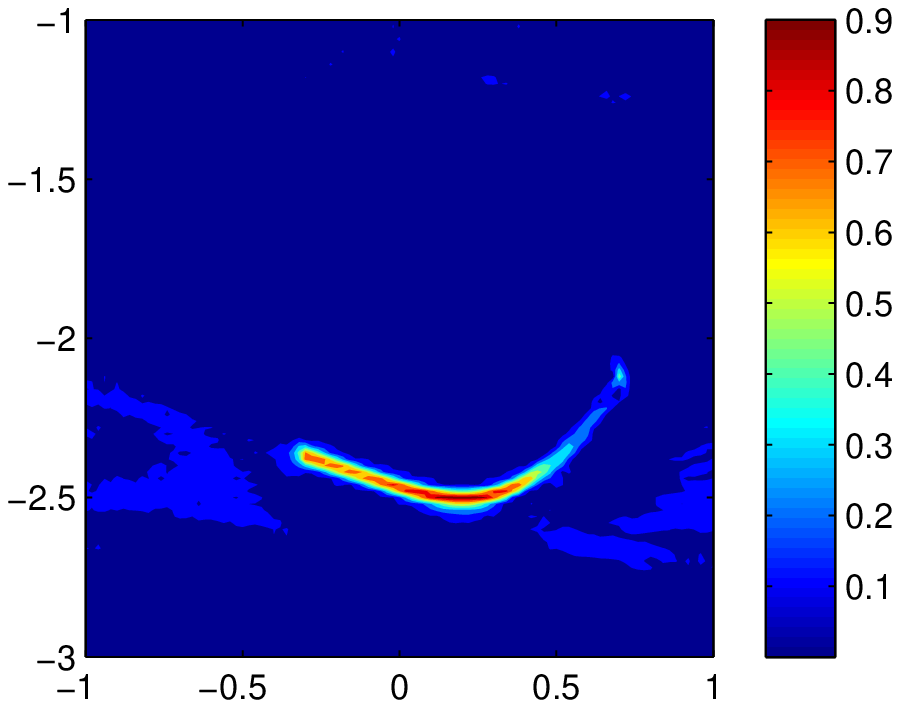}}\\
\subfigure[$\Gamma_{\mbox{\tiny M}}$ with $\eps_1=\eps_2=5$] {\label{MultiEps-1a}\includegraphics[width=0.49\textwidth]{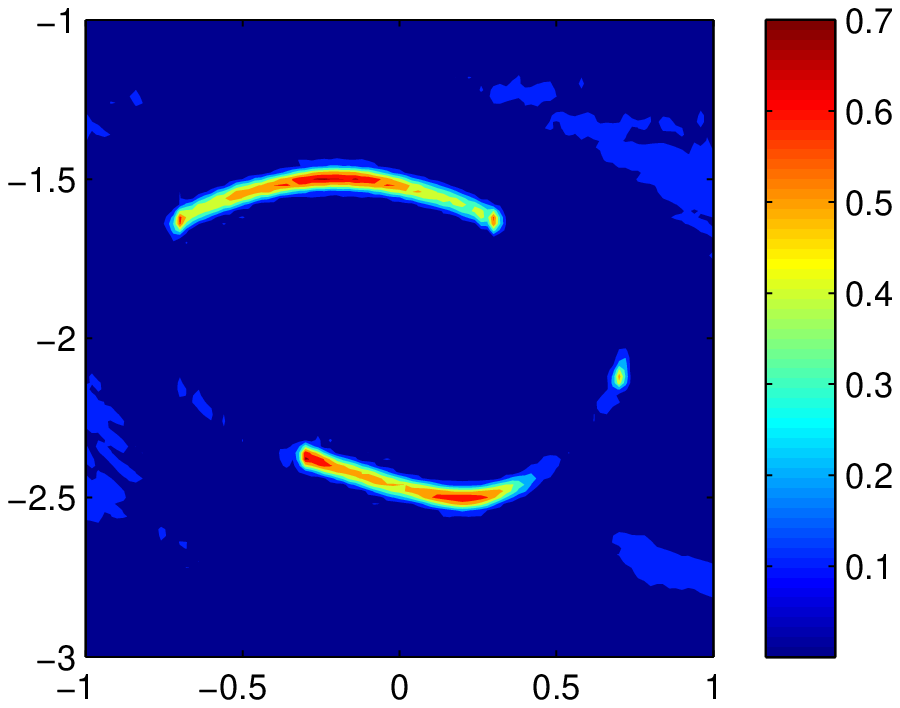}}
\subfigure[$\Gamma_{\mbox{\tiny M}}$ with $\eps_1=10$ and $\eps_2=5$] {\label{MultiEps-1b}\includegraphics[width=0.49\textwidth]{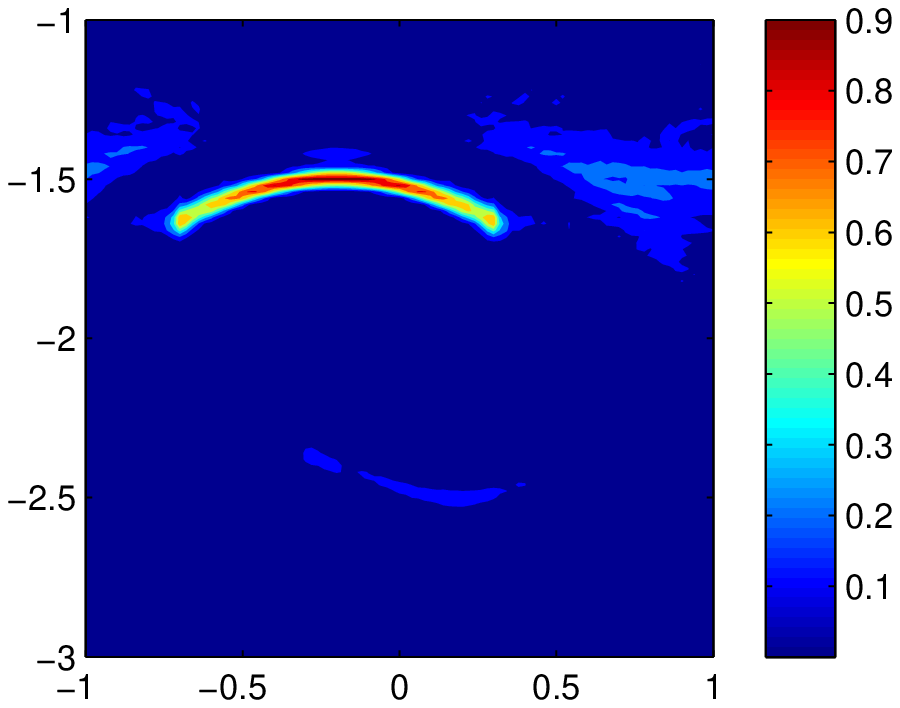}}
\caption{\label{GammaEps-1}($\eps_+>\eps_-$ case) Map of $W(x)$ for $\Gamma_1$, $\Gamma_2$ and $\Gamma_{\mbox{\tiny M}}$.}
\end{center}
\end{figure}

In this example, we choose the values $\eps_+=5$ and $\eps_-=4$. Thus, the number, $N_+$, of transmitted waves is smaller than the one, $N$, of directions of incidence. Based on the test configuration in Table \ref{Configuration1}, imaging results are depicted in Fig. \ref{GammaEps-1}. A good imaging resolution appears to be reached when the thin inclusion has a small constant curvature, e.g., Fig. \ref{SingleEps-1a}, but this resolution is poorer when it is of large curvature, e.g., Fig. \ref{SingleEps-1b}. Nevertheless, we can still conclude that a thin inclusion has been successfully retrieved.

The proposed algorithm could be applied directly to a target consisting of several well-separated thin inclusions with same thickness $h$. Let us skip the derivation herein and only provide imaging results. Those are depicted in Fig. \ref{GammaEps-1} with $\Gamma_1$ and $\Gamma_2$, the respective permittivities being $\eps_1$ and $\eps_2$. We notice that, if an inclusion has a much smaller value of permittivity than the other, it does not significantly affect the scattering matrix (as expected) and consequently, it cannot be retrieved via the proposed algorithm, refer to Fig. \ref{MultiEps-1b} (see \cite[Section 4.5]{PL2} also).

\subsubsection{$\eps_+<\eps_-$ case}
In contrast with the previous example, let us consider a more practical situation where the upper half-space (air) is the least refractive. In this case, we let $\eps_+=1$ and $\eps_-=3$. Now, the number of directions of transmitted waves, $N_+$, is the same as the one of directions of incidence, $N$.

\begin{figure}
\begin{center}
\subfigure[$\Gamma_1$ with $\eps_1=5$]{\label{SingleEps-2a}\includegraphics[width=0.49\textwidth]{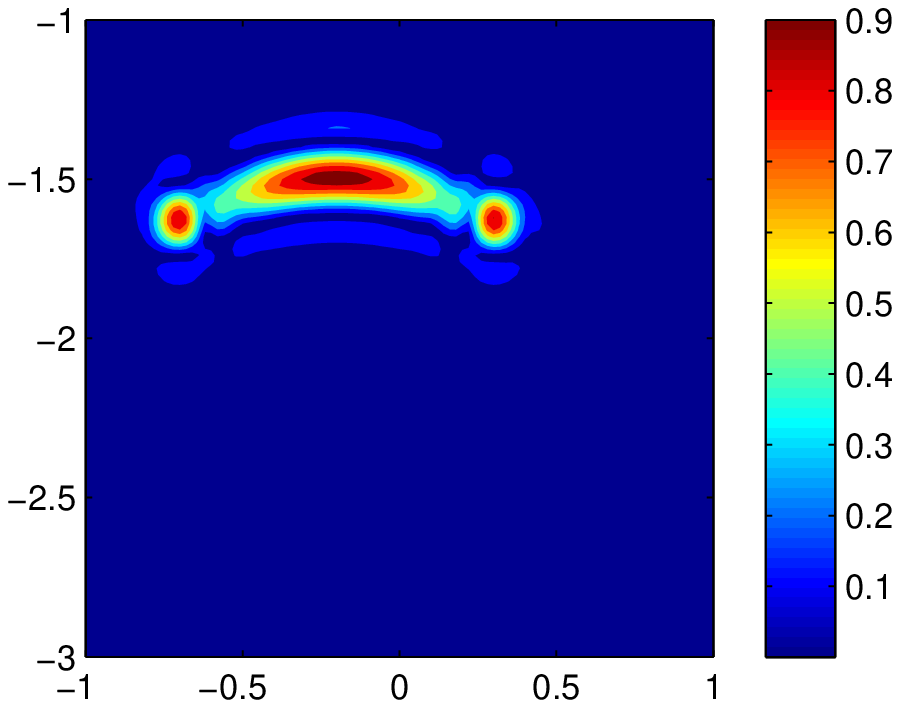}}
\subfigure[$\Gamma_2$ with $\eps_2=5$]{\label{SingleEps-2b}\includegraphics[width=0.49\textwidth]{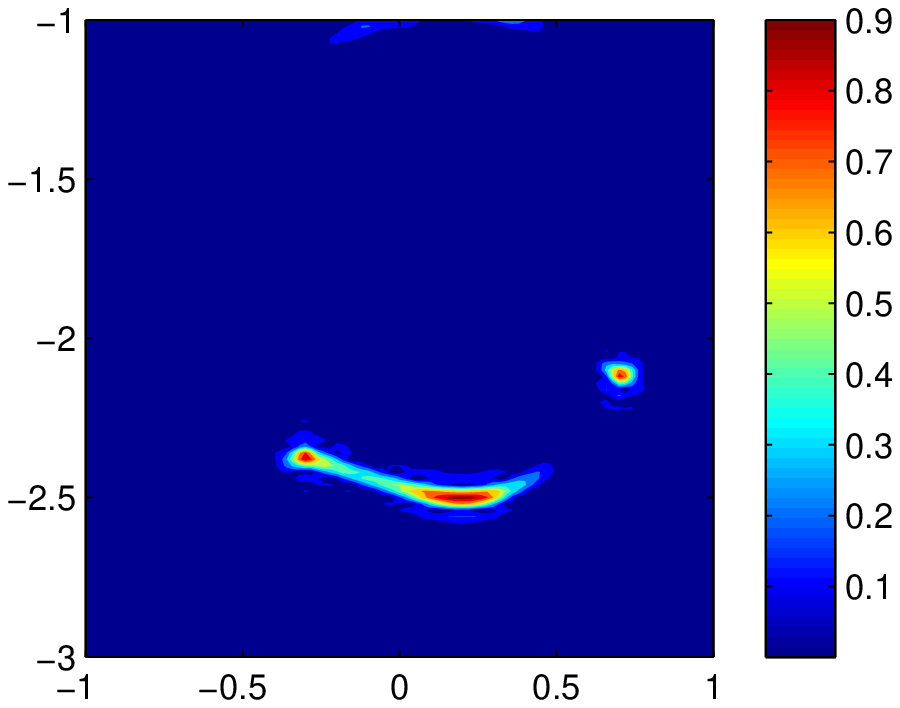}}\\
\subfigure[$\Gamma_{\mbox{\tiny M}}$ with $\eps_1=\eps_2=5$] {\label{MultiEps-2a}\includegraphics[width=0.49\textwidth]{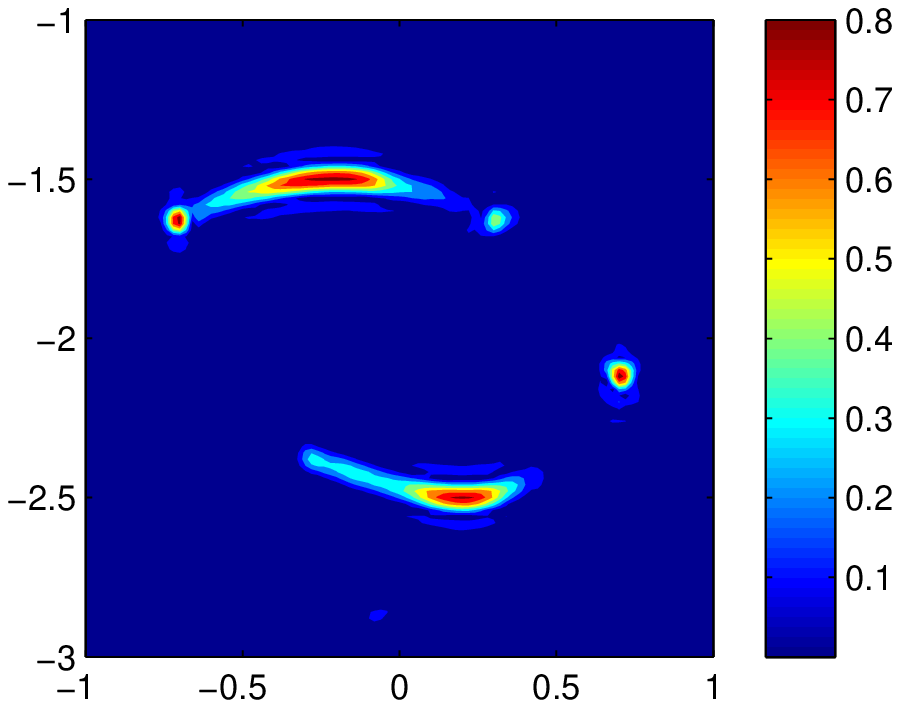}}
\subfigure[$\Gamma_{\mbox{\tiny M}}$ with $\eps_1=10$ and $\eps_2=5$] {\label{MultiEps-2b}\includegraphics[width=0.49\textwidth]{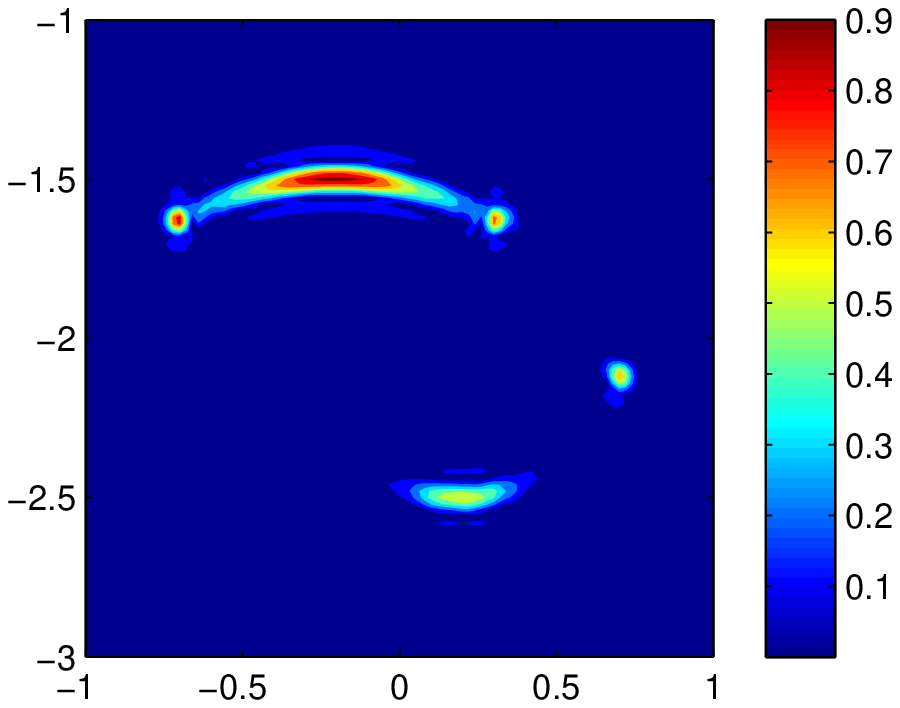}}
\caption{\label{GammaEps-2}($\eps_+<\eps_-$ case) Map of $W(x)$ for $\Gamma_1$, $\Gamma_2$ and $\Gamma_{\mbox{\tiny M}}$.}
\end{center}
\end{figure}

Figure \ref{GammaEps-2} shows the result based on the test configuration in Table \ref{Configuration1}. Similarly with the previous example, a thin inclusion of small constant curvature is well retrieved. However, poor results are observed when the inclusion is of large curvature. It is interesting to observe that the location of the end-points of $\Gamma_1$ and $\Gamma_2$ is well identified however. That is, connecting them by a straight line, it should provide a good initial guess for an iterative solution algorithm. Similarly with the Fig. \ref{MultiEps-1b}, whenever an inclusion has a significantly smaller value of permittivity than the other, it cannot be successfully retrieved via the proposed algorithm, refer to Fig. \ref{MultiEps-2b}.

\subsection{Permeability contrast case: $\eps_T=\eps_-=\eps_+$ and $\mu_T\ne\mu_-$}
Similarly with the previous case, two different situations of interest are considered based on the test configuration in Table \ref{Configuration2}.

\begin{table}
\begin{center}
\begin{tabular}{c|c|c|c|c|c}
\hline inclusion&$N$&$N_+$&$F$&frequency range&search domain\\
\hline\hline
$\Gamma_1$&$36$&$28$&$32$&$\omega\in[\frac{2\pi}{0.4},\frac{2\pi}{0.2}]$&$\Omega=[-1,1]\times[-3,-1]$\\
$\Gamma_2$&$42$&$36$&$36$&$\omega\in[\frac{2\pi}{0.3},\frac{2\pi}{0.2}]$&$\Omega=[-1,1]\times[-3,-1]$\\
$\Gamma_{\mbox{\tiny M}}$&$50$&$40$&$42$&$\omega\in[\frac{2\pi}{0.2},\frac{2\pi}{0.1}]$&$\Omega=[-1,1]\times[-3,-1]$\\
\hline
\end{tabular}
\caption{\label{Configuration2}(Permeability contrast case) Test configuration for $\Gamma_1$, $\Gamma_2$ and $\Gamma_{\mbox{\tiny M}}$.}
\end{center}
\end{table}

\subsubsection{$\mu_+>\mu_-$ case}
\begin{figure}
\begin{center}
\subfigure[$\Gamma_1$ with $\mu_1=5$]{\label{GammaMu-1a}\includegraphics[width=0.49\textwidth]{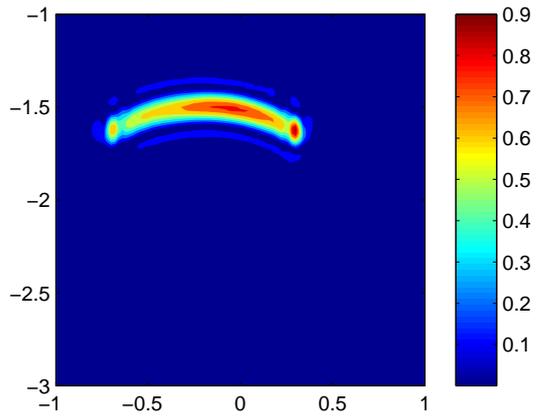}}
\subfigure[$\Gamma_2$ with $\mu_2=5$]{\label{GammaMu-1b}\includegraphics[width=0.49\textwidth]{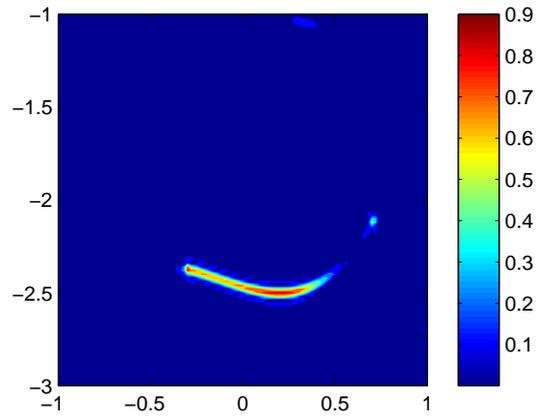}}\\
\subfigure[$\Gamma_{\mbox{\tiny M}}$ with $\mu_1=\mu_2=5$]{\label{MultiMu-1a}\includegraphics[width=0.49\textwidth]{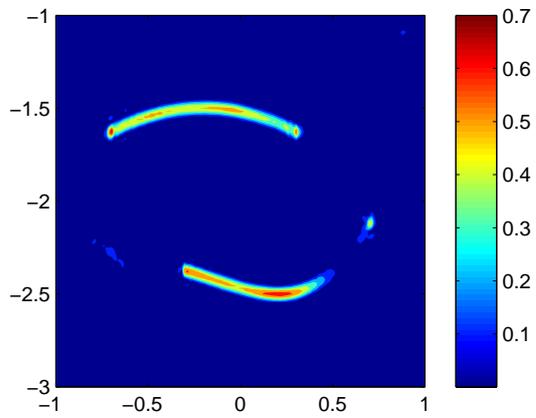}}
\subfigure[$\Gamma_{\mbox{\tiny M}}$ with $\mu_1=10$ and $\mu_2=5$]{\label{MultiMu-1b}\includegraphics[width=0.49\textwidth]{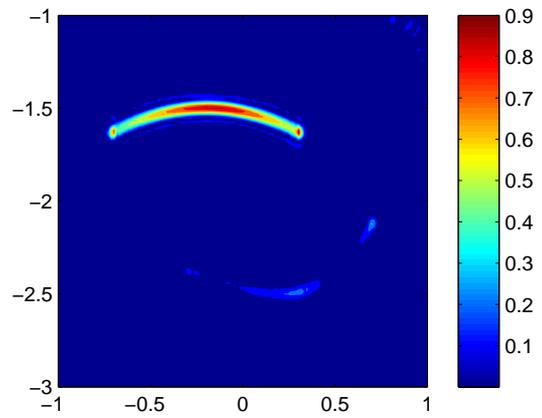}}
\caption{\label{GammaMu-1}($\mu_+>\mu_-$ case) Map of $W(x)$ for $\Gamma_1$, $\Gamma_2$ and $\Gamma_{\mbox{\tiny M}}$.}
\end{center}
\end{figure}

In this example, we choose the values $\mu_+=5$ and $\mu_-=4$. Thus, we have a number, $N_+$, of transmitted waves which is smaller than the one, $N$, of directions of incidence. $b$ is set to c$b=(0,1)$ (see Remark \ref{Rem}). Based on the test configuration in Table \ref{Configuration2}, imaging results are depicted in Fig. \ref{GammaMu-1}. Similarly with the permittivity contrast case, a good imaging resolution is attained when the thin inclusion has a small constant curvature but a poorer resolution when it is of large curvature. In the result corresponding to multiple inclusions, similarly with what happens in the permittivity contrast case, we easily notice that if an inclusion has a much smaller value of permeability than the other, it cannot be retrieved via the proposed algorithm, refer to Fig. \ref{MultiMu-1b}.

\subsubsection{$\mu_+<\mu_-$ case}
In contrast with the previous example, let us consider that the upper half space is the least refractive. In this case, we set $\mu_+=1$ and $\mu_-=3$.

\begin{figure}
\begin{center}
\subfigure[$\Gamma_1$ with $\mu_1=5$]{\label{GammaMu-2a}\includegraphics[width=0.49\textwidth]{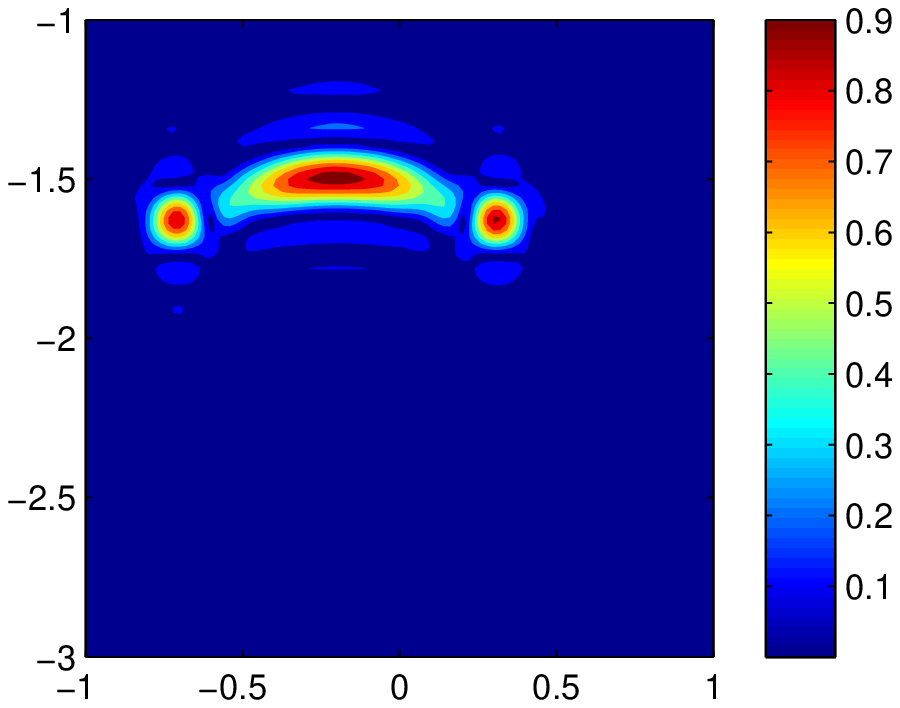}}
\subfigure[$\Gamma_2$ with $\mu_2=5$]{\label{GammaMu-2b}\includegraphics[width=0.49\textwidth]{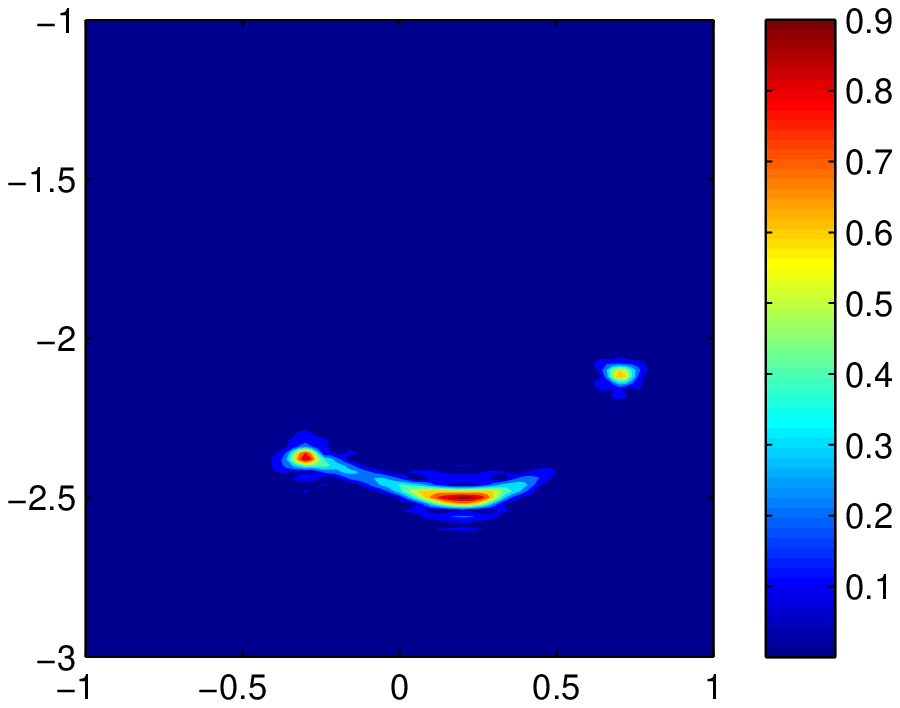}}\\
\subfigure[$\Gamma_{\mbox{\tiny M}}$ with $\mu_1=\mu_2=5$]{\label{MultiMu-2a}\includegraphics[width=0.49\textwidth]{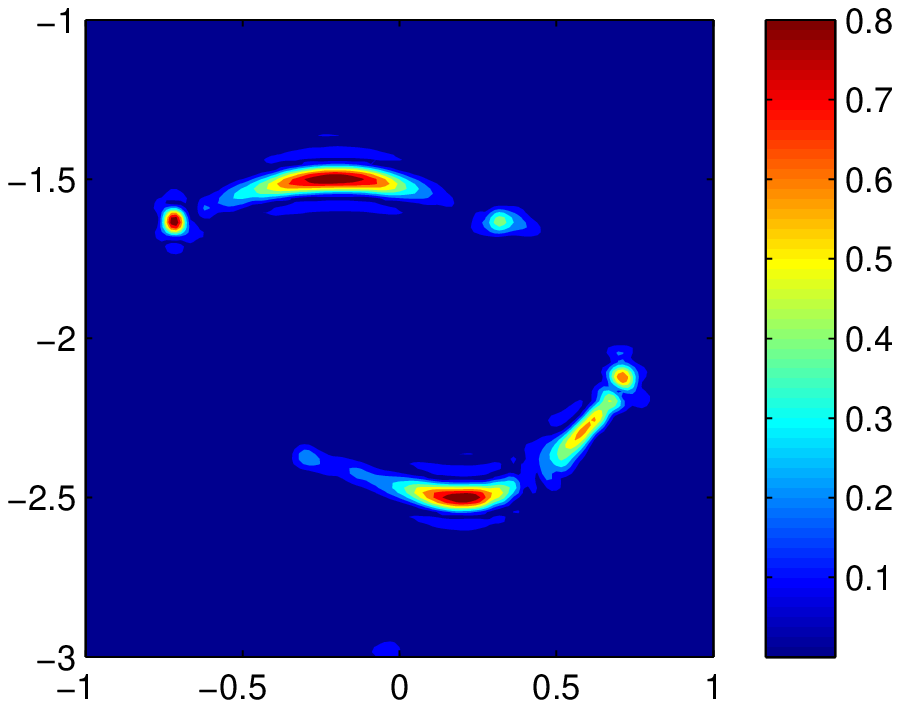}}
\subfigure[$\Gamma_{\mbox{\tiny M}}$ with $\mu_1=10$ and $\mu_2=5$]{\label{MultiMu-2b}\includegraphics[width=0.49\textwidth]{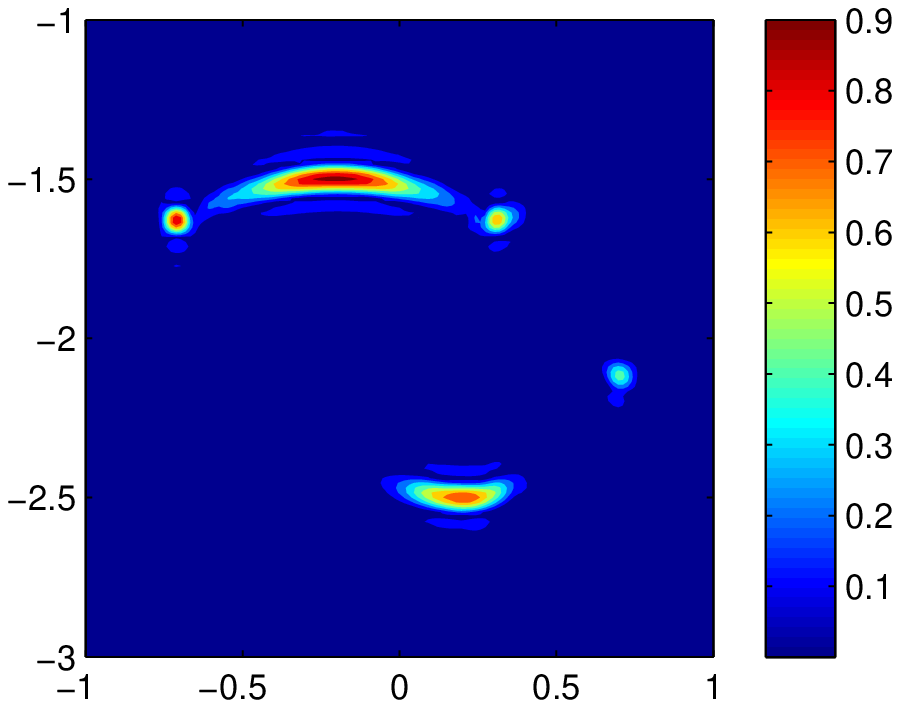}}
\caption{\label{GammaMu-2}($\mu_+<\mu_-$ case) Map of $W(x)$ for $\Gamma_1$, $\Gamma_2$ and $\Gamma_{\mbox{\tiny M}}$.}
\end{center}
\end{figure}

Results are shown in Fig. \ref{GammaMu-2} based on the test configuration in Table \ref{Configuration2}. Similarly with the $\eps_+<\eps_-$ case case (Fig. \ref{GammaEps-2}), the location of the end-points of the thin inclusion is well identified. As for the imaging of multiple inclusions, tough we cannot reach a proper result, a good initial guess is made possible.

\subsection{Permittivity and permeability contrast case: $\eps_T\ne\eps_-$ and $\mu_T\ne\mu_-$}
Similarly with the previous cases, two different situations of interest are considered based on the test configuration given in Table \ref{Configuration3}.

\begin{table}
\begin{center}
\begin{tabular}{c|c|c|c|c|c}
\hline inclusion&$N$&$N_+$&$F$&frequency range&search domain\\
\hline\hline
$\Gamma_1$&$40$&$32$&$36$&$\omega\in[\frac{2\pi}{0.4},\frac{2\pi}{0.2}]$&$\Omega=[-1,1]\times[-3,-1]$\\
$\Gamma_2$&$48$&$40$&$40$&$\omega\in[\frac{2\pi}{0.3},\frac{2\pi}{0.2}]$&$\Omega=[-1,1]\times[-3,-1]$\\
$\Gamma_{\mbox{\tiny M}}$&$60$&$50$&$48$&$\omega\in[\frac{2\pi}{0.2},\frac{2\pi}{0.1}]$&$\Omega=[-1,1]\times[-3,-1]$\\
\hline
\end{tabular}
\caption{\label{Configuration3}(Both permittivity and permeability contrast case) Test configuration for $\Gamma_1$, $\Gamma_2$ and $\Gamma_{\mbox{\tiny M}}$.}
\end{center}
\end{table}

\subsubsection{$\eps_+>\eps_-$ and $\mu_+>\mu_-$ case}
\begin{figure}
\begin{center}
\subfigure[$\Gamma_1$ with $\eps_1=\mu_1=5$]{\label{GammaEpsMu-1a}\includegraphics[width=0.49\textwidth]{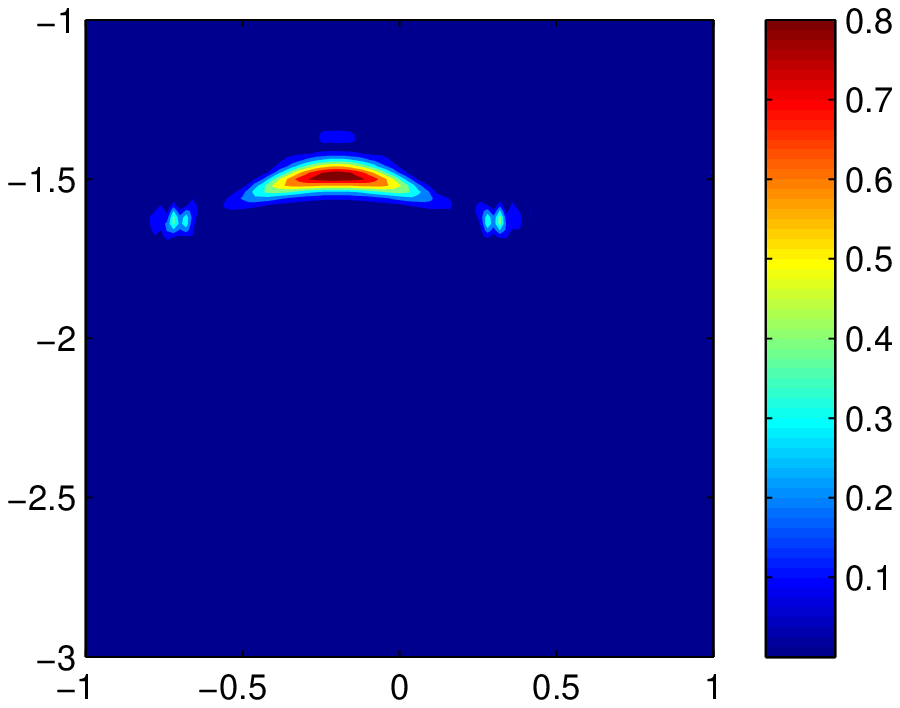}}
\subfigure[$\Gamma_2$ with $\eps_2=\mu_2=5$]{\label{GammaEpsMu-1b}\includegraphics[width=0.49\textwidth]{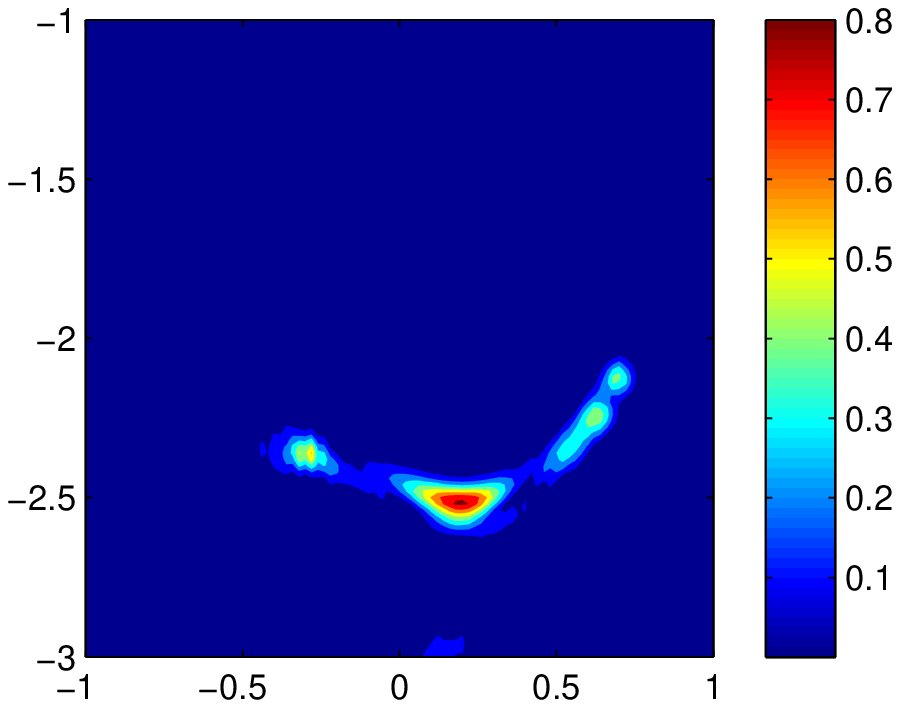}}\\
\subfigure[$\Gamma_{\mbox{\tiny M}}$ with $\eps_1=\eps_2=\mu_1=\mu_2=5$]{\label{MultiEpsMu-1a}\includegraphics[width=0.49\textwidth]{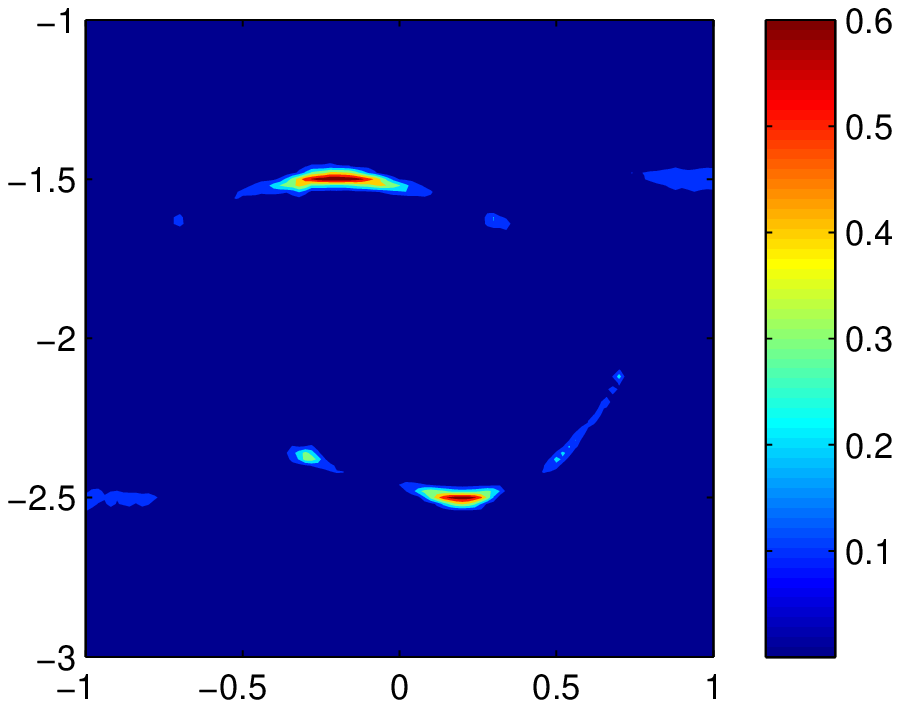}}
\subfigure[$\Gamma_{\mbox{\tiny M}}$ with $\eps_1=\mu_1=10$ and $\eps_2=\mu_2=5$]{\label{MultiEpsMu-1b}\includegraphics[width=0.49\textwidth]{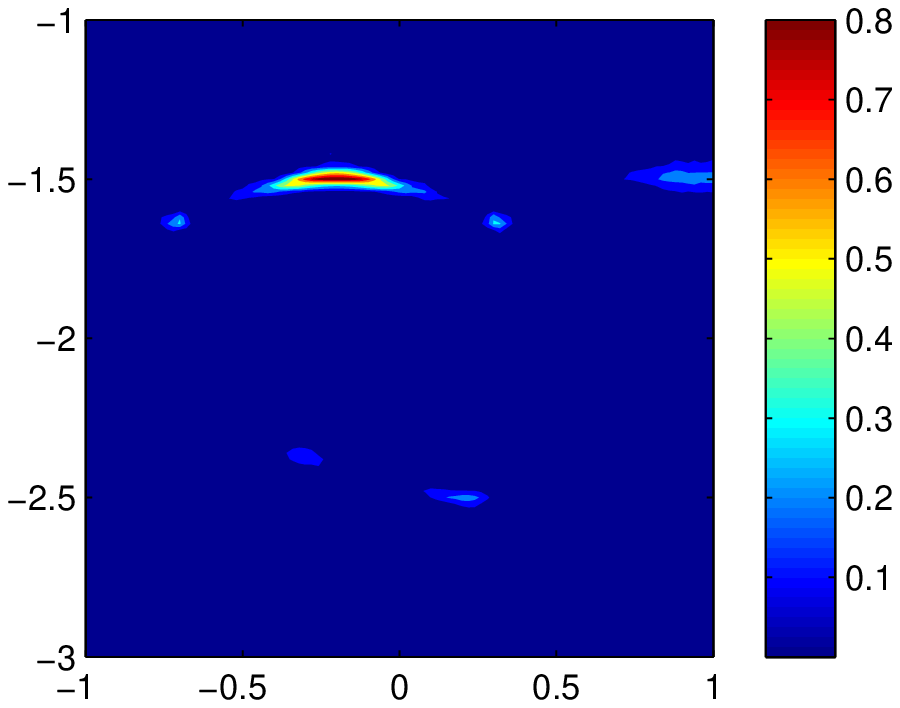}}
\caption{\label{GammaEpsMu-1}($\eps_+>\eps_-$ and $\mu_+>\mu_-$ case) Map of $W(x)$ for $\Gamma_1$, $\Gamma_2$ and $\Gamma_{\mbox{\tiny M}}$.}
\end{center}
\end{figure}

In this case, the values $\eps_+=5$, $\mu_+=5$, $\eps_-=4$ and $\mu_-=4$ are chosen, i.e., $c$ is set to $c=(1,b)=(1,0,1)$ (see Remark \ref{Rem}). Based on the test configuration in Table \ref{Configuration3}, imaging results are depicted in Fig. \ref{GammaEpsMu-1}. In opposition with the previous cases, a poorer resolution is now achieved. However, at least, we can approximate the shape of the thin inclusions unless one has a much smaller value of permittivity and permeability than the other.

\subsubsection{$\eps_+<\eps_-$ and $\mu_+<\mu_-$ case}
\begin{figure}
\begin{center}
\subfigure[$\Gamma_1$ with $\eps_1=\mu_1=5$]{\label{GammaEpsMu-2a}\includegraphics[width=0.49\textwidth]{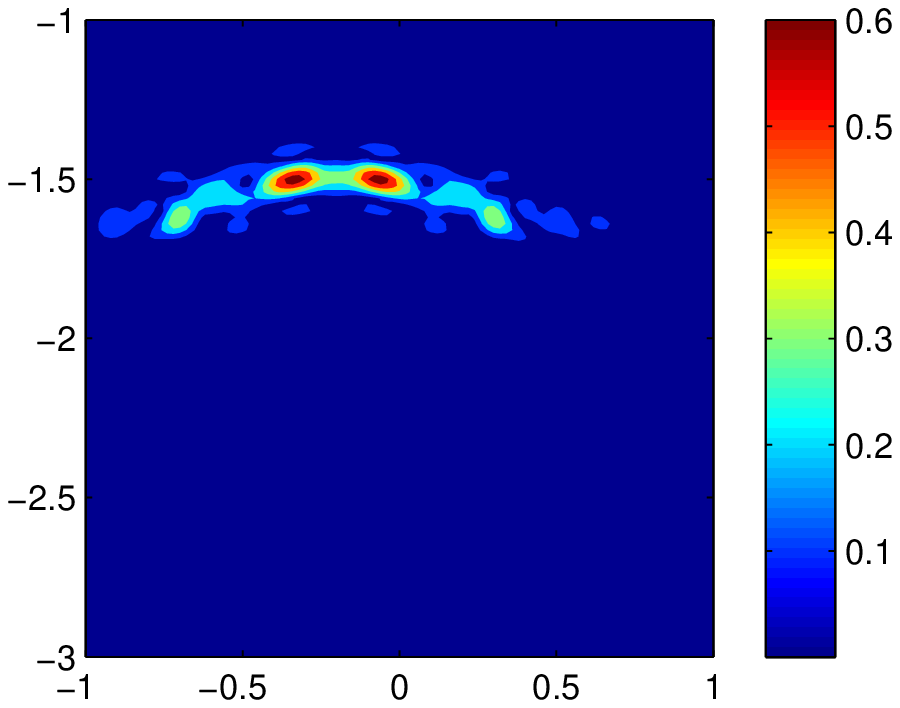}}
\subfigure[$\Gamma_2$ with $\eps_2=\mu_2=5$]{\label{GammaEpsMu-2b}\includegraphics[width=0.49\textwidth]{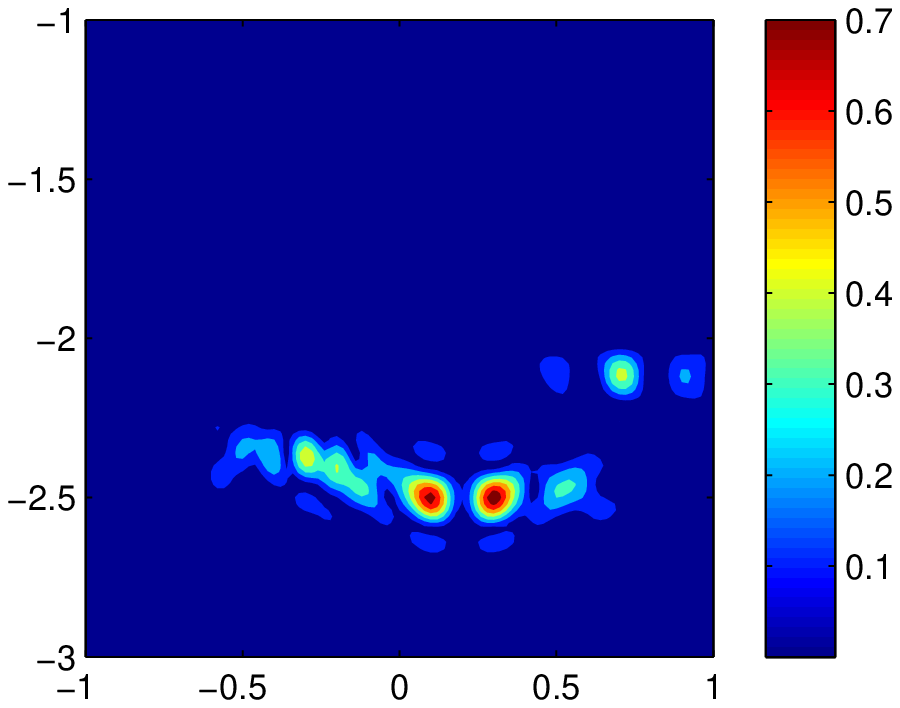}}\\
\subfigure[$\Gamma_{\mbox{\tiny M}}$ with $\eps_1=\eps_2=\mu_1=\mu_2=5$]{\label{MultiEpsMu-1c}\includegraphics[width=0.49\textwidth]{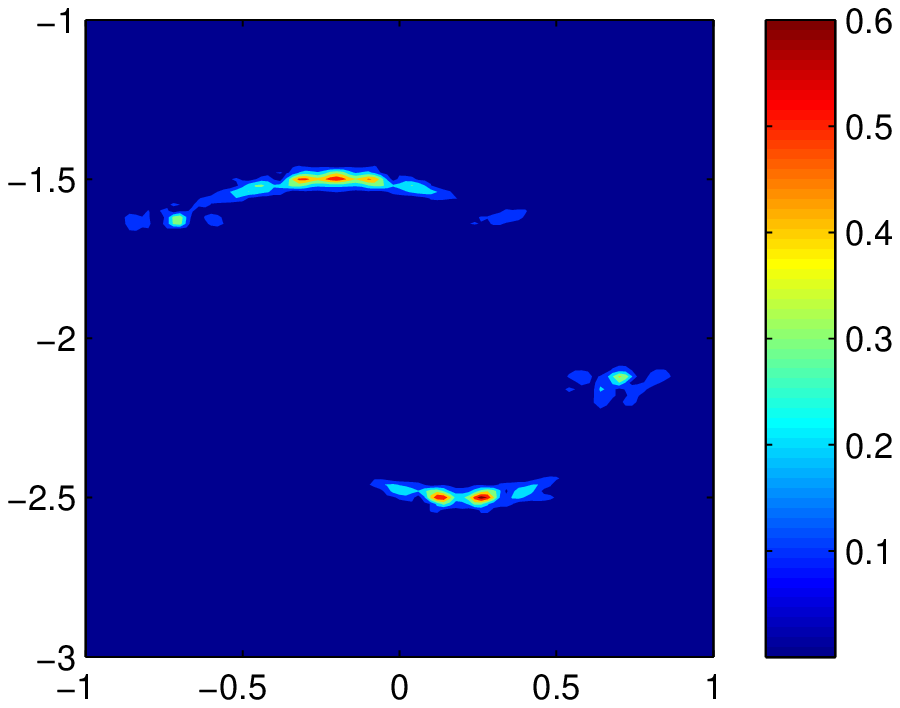}}
\subfigure[$\Gamma_{\mbox{\tiny M}}$ with $\eps_1=\mu_1=10$ and $\eps_2=\mu_2=5$]{\label{MultiEpsMu-1d}\includegraphics[width=0.49\textwidth]{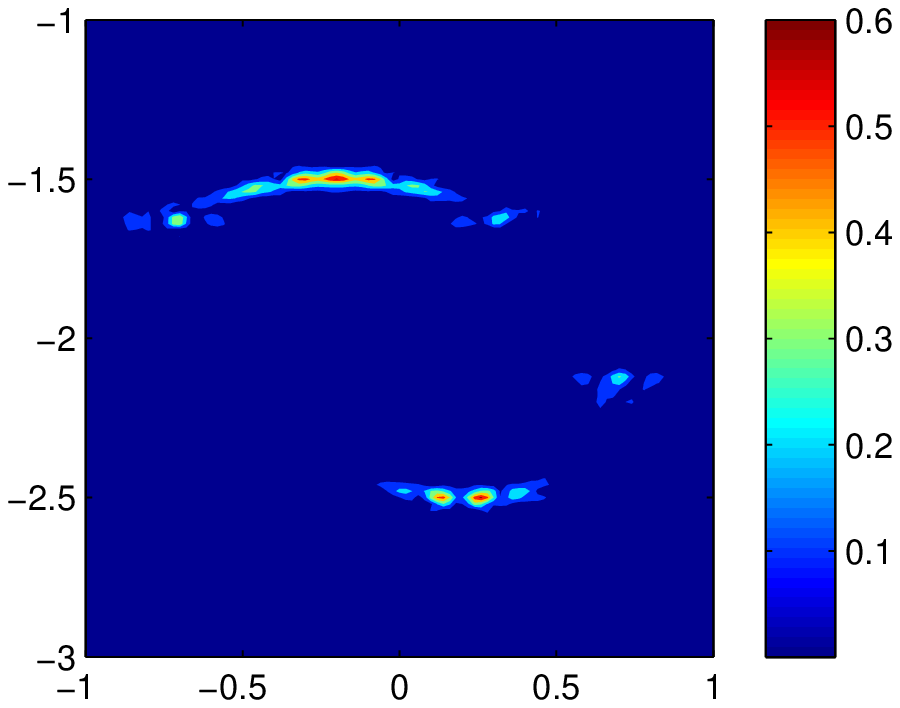}}
\caption{\label{GammaEpsMu-2}($\eps_+<\eps_-$ and $\mu_+<\mu_-$ case) Map of $W(x)$ for $\Gamma_1$, $\Gamma_2$ and $\Gamma_{\mbox{\tiny M}}$.}
\end{center}
\end{figure}

Now, let us consider that the upper half space is the least refractive. In this case, we let $\eps_+=1$, $\mu_+=1$, $\eps_-=3$ and $\mu_-=3$. Based on the test configuration in Table \ref{Configuration3}, imaging results are depicted in Figs. \ref{GammaEpsMu-2}. Although poor results are found when the inclusion is of arbitrary shape, opposite to the previous examples, we still recognize a part of the thin inclusion $\Gamma_2$ (neighborhood of point $(0.3,-2.5)$) has a much smaller value of permittivity and permeability than the other, refer to Fig. \ref{MultiEpsMu-1d}.

\section{Conclusion}\label{Sec6}
A non-iterative imaging algorithm operated at several time-harmonic frequencies has been proposed in order to determine the unknown support of a thin penetrable electromagnetic inclusion buried within a half space. The main idea comes from the fact that the collected Multi-Static Response (MSR) matrix data can be modeled via a rigorous asymptotic formulation for a limited range of incident and observation directions.

Thanks to various numerical simulations, it is shown that the proposed algorithm is both effective and robust with respect to random noise. Moreover, it can be easily applied to multiple inclusions. Nevertheless, some improvements are still required, e.g., when the sought inclusion is of large curvature or has a significantly smaller value of permittivity (or permeability) than, say, another nearby. In addition, in order to achieve a proper imaging of arbitrarily shaped thin inclusions in the case of permeable, or dielectric and permeable material, the choice of the testing vector functions $b$ and $c$ still requires further mathematical investigation.

It is worthwhile emphasizing that such results are obtained at low computational cost but that they do not guarantee complete shaping of the inclusions. However, they could provide a good initial guess of a level-set evolution \cite{ADIM,DL,PL4} or of any other standard iterative algorithm \cite{DEKPS}.

Finally, we have been considering a two-dimensional problem. The strategy which is suggested, e.g., mathematical treatment of the asymptotic formula, imaging algorithm, etc., could be extended to the three-dimensional problem, refer to \cite{AILP,GLCP,IGLP} for related work.

\section*{Acknowledgement}
W.-K. Park would like to acknowledge Professor Karl Kunisch for his hospitality during the post doc. period at Institute for Mathematics and Scientific Computing, University of Graz.

\end{document}